\documentclass[12pt] {article}

\usepackage{amssymb,amsmath,amsthm}
\usepackage[round]{natbib} % gives options like citet, citep
\usepackage{color}
\usepackage{hyperref}
\usepackage{graphicx}%[dvips]

\newcommand{\beq}[1]{  \begin{equation} \label{#1} }
\newcommand{\eeq}{     \end{equation}}
\newcommand{\bal}[1]{\begin{align} \label{#1} }

\theoremstyle{plain}
\newtheorem{thm}{Theorem}
\newtheorem{lem}{Lemma}  %\theoremstyle{Lemma}
\theoremstyle{remark}
\newtheorem{rmk}{Remark}
\newtheorem{examp}[thm]{Example}

\newcommand{\rf}[1]{(\ref{#1})}
\def\bd#1{\mbox{\boldmath$\displaystyle\mathbf{#1}$} }
\def\tr{\operatorname{tr}}
\def\Div{\operatorname{Div}}

\def\singlespacing{\baselineskip=13pt}  

\begin{document} %%%%%%%%%%%%%%%%%%%%%%%%%%%%%%%%%%%%%%%%%%%%%%%%%%%%%%%%%%%%%%%%%%%%%%%%%

\pagestyle{myheadings}\markright{{\sc Degenerate  nonlinear elastic  waves}  ~~~~~~\today}
\singlespacing%\doublespacing

\title{\color{blue} Degenerate weakly nonlinear elastic plane waves}
\author{W{\l}odzimierz Doma\'nski\footnote{Military University of Technology, Faculty of Cybernetics,
Institute of Mathematics and Cryptology, Gen. S. Kaliskiego 2,
00-908 Warsaw 49, Poland, domanski.wlodek@gmail.com} $\,$ and Andrew
N. Norris\footnote{Rutgers University, Department of Mechanical and
Aerospace Engineering, 98 Brett Road, Piscataway, NJ  08854-8058,
norris@rutgers.edu}}\maketitle
\thanks{\centerline {\it Dedicated to Philippe Boulanger on the occasion of his sixtieth
birthday.}}

\begin{abstract}

Weakly nonlinear plane waves are considered in hyperelastic
crystals. Evolution equations are derived at a quadratically
nonlinear level for the amplitudes of quasi-longitudinal and
quasi-transverse waves propagating in arbitrary anisotropic media.
The form of the  equations obtained depends upon the direction of
propagation relative to the crystal axes.  A single equation is
found for all propagation directions for quasi-longitudinal waves,
but a pair of coupled equations occurs for quasi-transverse waves
propagating along directions of degeneracy, or acoustic axes. The
coupled equations involve four material parameters but they simplify
if the wave propagates  along an axis of  material symmetry.  Thus,
only two parameters arise for propagation along an axis of two-fold
symmetry, and one for a three-fold axis. The transverse wave
equations decouple if the axis is four-fold or higher.  In the
absence of a symmetry axis it is possible that the evolution
equations {of the quasi-transverse waves decouple if the third order
elastic moduli satisfy a certain identity.}  The theoretical results are illustrated with
explicit examples.

%A third possibility arises when the second order nonlinearity coefficient vanishes, as for transverse waves in isotropic materials.  The results together show that three distinct types of nonlinear equations governing the amplitude evolution of weakly nonlinear elastic plane waves in generally anisotropic elastic solids.

\end{abstract}
\section{Introduction}\label{sec1}

We characterize and analyze degenerate weakly nonlinear elastic
waves in crystals. The degeneracy considered here arises from the
existence of acoustic axes in elastic materials. Acoustic axes are
directions for which the phase velocities of at least two waves
coincide. For classical elasticity this phenomenon  {typically
occurs} for transverse or quasi-transverse waves. In the
mathematical literature the coincidence  of wave speeds is called a
loss of strict hyperbolicity. Conditions for the existence of
acoustic axes were first derived by \citet{khatkevich1962}, and a
useful review of the topic is given by \citet{fed}. Recent
developments can be found in e.g. \cite{bh98,mozhaev2001,Norris04b}.
{Analysis and properties of nonlinear elastic
waves propagating along acoustic axes were discussed in
\cite{Shuvalov01}, see also the book of \cite{Lyamov83}.}

The existence of acoustic axes, i.e. the loss of strict
hyperbolicity, is typically accompanied by the local loss of genuine
nonlinearity, that is, vanishing of the scalar product of the
gradient of the phase velocity with the corresponding polarization
vector, all evaluated at the origin. The local loss of genuine
nonlinearity implies the presence of weaker than quadratic (e.g.
cubic) nonlinearities in the decoupled evolution equations for
degenerate weakly nonlinear (quasi-)transverse waves. This happens
e.g. for shear elastic plane waves in an isotropic material as well
as in a cubic crystal e.g. for [1 0 0] or [1 1 0] directions (see
\citet{Domanski00}). However, coupled quadratically nonlinear
evolution equations do occur for transverse elastic waves. Although
this is not possible in isotropic materials \citep{Goldberg60}, such
couplings can manifest themselves for special directions in
crystals, for instance,  for propagation in the [1 1 1] direction in
a cubic crystal (see \citet{Domanski00}). These special directions
are the acoustic axes.

In this paper we clarify what kind of coupling is possible according
to the type of a symmetry axis and we demonstrate how many constants
are needed to describe the coupling of pairs of (quasi-)transverse
waves in a particular case. It turns out that we need two constants
for 2-fold symmetry axis, and only one constant for 3-fold axis.
Moreover we also prove that there  cannot be a quadratically
nonlinear coupling for shear wave equations if the symmetry axis is
4-fold or 6-fold. In the absence of symmetry, the propagation of
shear waves along an acoustic axis depends upon four constants
governing the nonlinear terms in the coupled equations.  It is
possible that  a pair of evolution equations decouples if the four
constants satisfy certain special relations, which are derived here.
We illustrate these general statements with some examples of
particular elastic materials.  Some of these results appeared
previously in an abbreviated form \citep{Domanski08}.
%The full details and derivations are provided here.

The paper is organized as follows.  In Section \ref{sec2} we present
the model of nonlinear elastodynamics and its  constitutive
assumptions and we demonstrate the reduction of the governing
equations to a  quasilinear plane wave  system. Section \ref{sec3}
contains a  presentation of the method of Weakly Nonlinear Geometric
Optics (WNGO) and its applications to quasi-longitudinal and
quasi-transverse waves. Special attention is devoted to the
degenerate case of pairs of quasi-transverse waves propagating along
acoustic axes.  Coupled evolution equations are derived for two-fold
and three-fold symmetry axes.  The analysis of the coupled evolution
equations is discussed in Section \ref{sec4} under conditions of
symmetry about the propagation direction. Explicit examples which
illustrate some of these theoretical results are provided for cubic
crystals. Section \ref{sec5} considers the special case of
{quasi-transverse} waves propagating along an acoustic axis in the
absence of material symmetry, and derives a condition both necessary
and sufficient that the evolution equations decouple.   Some
concluding remarks are offered at the end of the paper.

\section{Preliminaries}\label{sec2}

\subsection{Basic equations}

The equation of motion of a continuum written in material
(Lagrangian) coordinates is, in the absence of body forces, \beq{01}
\rho_0 \frac{\partial^2 {\bd u}} {\partial t^2}= \Div {\bd T}. \eeq
Here $\rho_0$ is the mass density in the reference configuration,
$\bd u$ is the particle displacement, $\bd T$ is the first
Piola-Kirchhoff stress tensor, and $\Div$ denotes the divergence
operator with respect to the material coordinates ${\bd X}$. In the
hyperelastic medium  there exists  a stored energy density per unit
volume in the reference configuration, denoted by $W(\bd F) $, such
that \beq{02} {\bd T} = \frac{\partial W}{\partial {\bd F}}, \eeq
where the deformation gradient  is \beq{03} {\bd F} = {\bd I} +
\nabla {\bd u} \eeq  
{with  ${\bd I}$ the identity tensor.}

\subsection{First order system}

It is convenient for our purposes  to write the equation of motion as
a first order system of partial differential equations. To this aim we introduce the particle velocity
\beq{05}
{\bd v} = \frac{\partial {\bd u}} {\partial t},
\eeq
and instead of  (\ref{01}) we write the  system of equations
\begin{subequations}\label{06ab}
\bal{06}
\rho_0 \frac{\partial {\bd v}} {\partial t}&= \Div
 {\bd T},
\\
 \frac{\partial  {\bd F}} {\partial t}&=
\nabla  {\bd v}.
\label{07}
\end{align}
\end{subequations}
Equation (\ref{07}) follows from the comparison of the
time derivative of (\ref{03}) with the space gradient of (\ref{05}),
assuming that the displacement vector ${\bd u}$  is at least twice
continuously differentiable.
We introduce the following notation for the displacement gradient
\beq{040}
{\bd M}  \equiv \nabla {\bd u}  .
\eeq
Then, using (\ref{02}) and (\ref{03}) in the form ${\bd F} ={\bd I} +{\bd M} $ we can express the system  (\ref{06ab})
 as
\begin{subequations}\label{08ab}
\bal{08}
\rho_0 \frac{\partial {\bd v}} {\partial t}&= \Div\, \frac{\partial W}{\partial {\bd M}},
\\
 \frac{\partial  {\bd M}} {\partial t}&=
\nabla  {\bd v}.
\label{09}
\end{align}
\end{subequations}

\subsection{Constitutive relations}

Components of vectors and other quantities will be referred to an
orthonormal basis $\{ {\bd e}_1,\, {\bd e}_2,\, {\bd e}_3\}$. The
strain energy in an arbitrary material is assumed to have the
following expansion for small strains: \beq{11}
 W = \frac{1}{2!}\,
 c_{abcd}\, E_{ab}\, E_{cd}
+  \frac{1}{3!}\,c_{abcdef}\, E_{ab}\, E_{cd}\, E_{ef} + \cdots ~,
\eeq
where $E_{ab}$ are the components of the left Cauchy-Green strain tensor
\beq{10}
{\bd E} = \frac12 ({\bd F}^t {\bd F} - {\bd I}) =
\frac12 ( {\bd M} + {\bd M}^t+ {\bd M}^t {\bd M} ),
\eeq
and  the summation convention on repeated subscripts $a,b,\cdots,
g,h$ is assumed. The symmetry of the strain implies the relations
\beq{12}
c_{abcd} = c_{cdab}  = c_{bacd} ,\qquad c_{abcdef} =
c_{cdabef}  = c_{cdefab}  =c_{bacdef},
\eeq
which imply that there are no more than 21 and  56 independent second and third order moduli, respectively. The second order moduli are assumed to be positive definite in the sense that $ c_{abcd} s_{ab} s_{cd} >0$ for all non-zero ${\bd s}={\bd s}^t$.

\citet{Hearmon53} provides a complete
enumeration of the third-order constants for all crystal classes.
  Equations \rf{02}, \rf{03}, \rf{11} and \rf{10} together imply that the Piola-Kirchhoff stress is
\beq{13}
T_{ab} %=  \frac{\partial W}{\partial M_{ab}}
=  c_{abcd}\,
M_{cd}
 + \frac{1}{2} N_{abcdef}\,M_{cd}M_{ef} + \frac{1}{6} N_{abcdefgh}\,
M_{cd}M_{ef}M_{gh} + \cdots ~,
\eeq
where
\beq{14}
N_{abcdef}  =  c_{abcdef} +  c_{abdf} \, \delta_{ce} +  c_{bfcd} \,
\delta_{ae} +  c_{bdef} \, \delta_{ac}\, .
\eeq
\citet[eq. (38.5)]{Thurston84} gives an expression for the higher order
coefficients $ N_{abcdefgh}$.  Note that $N_{abcdef}\ne N_{bacdef}
$, which implies that the non-symmetry of $\bf T$ is a second-order
effect.

For the remainder of the paper we take $\rho_0 = 1$ for simplicity.

%%%%%%%%%%%%%%%%%%%%%

\subsection{Plane waves}
Plane wave solutions are described by    displacement ${\bd u}$ that depends
upon  a single component of ${\bd X}$, say $x = {\bd X}\cdot{\bd n}$,
where ${\bd n}$ is the direction of propagation. The  displacement
gradient of \rf{040}
   reduces to
\beq{15}
{\bd M} %\equiv \nabla {\bd u}
= {\bd m}\otimes
{\bd n},
\eeq
where
\beq{153}
{\bd m} = \frac{\partial {\bd u}} {\partial x},
\eeq
is a  displacement gradient vector. Defining the energy function for plane deformation by
\beq{16}
V(\bd m) \equiv W(\bd m \otimes \bd n),
\eeq{}
we can write the system of  elastodynamic equations \rf{08ab}
for plane waves as
\begin{subequations}\label{17ab}
\bal{17}
 \frac{\partial {\bd v}} {\partial t}& =
\frac{\partial^2 V}{\partial {\bd m}^2}\cdot
\frac{\partial {\bd m}}{\partial x},
\\
  \frac{\partial  {\bd m}} {\partial t}&=
\frac{\partial {\bd v}}{\partial x} .
\label{18}
\end{align}
\end{subequations}
The above system can be expressed in a quasilinear form  as
%\footnote{From now on we assume that $\varrho_0 \equiv 1$ for simplicity.}
\beq{19}
\frac{\partial{\bd w}} {\partial t} +{\bd A}({\bd w})
\frac{\partial{\bd w}}{\partial x}= {\bd 0} ,
\eeq
where
\beq{20}
{\bd w} = \begin{pmatrix}
{\bd v} \\
{\bd m}
\end{pmatrix}
,
\qquad
{\bd A}
({\bd w}) =
- \begin{bmatrix}
{\bd 0} &
{\bd B} \\
{\bd I} &
{\bd 0}
\end{bmatrix}
.
\eeq
The $3\times 3$ matrix $\bd B$ is
\beq{21}
{\bd B} = {\bd \Lambda} + {\bd \Psi}{\bd m} + \frac12 {\bd \Pi}{\bd m}{\bd m}
+ \cdots\, ,
\eeq
where  ${\bd \Lambda}({\bd n})$, ${\bd \Psi}({\bd n})$ and ${\bd \Pi}({\bd n})$ are, in   components,
\begin{subequations}
\begin{align}
\Lambda_{ac}  &=c_{abcd}n_bn_d,
\\
\Psi_{ace}  &= N_{abcdef}  n_bn_dn_f,
\\
\Pi_{aceg}  &= N_{abcdefgh}n_bn_dn_fn_h,
\end{align}
\end{subequations}
or, in short,
\beq{221}
 B_{ac}  =c_{abcd}n_bn_d
 + N_{abcdef}  n_bn_dn_f \,m_e+\frac12
  N_{abcdefgh}n_bn_dn_fn_h\, m_em_g
+ \cdots\, .
\eeq

%\subsection{Eigenvalues of ${\bf A}(0)$}

The positive definite  property of the second order moduli  implies that
${\bd \Lambda}$, also known as the Christoffel or acoustical tensor, has spectral  form
\beq{231}
{\bd \Lambda} = \alpha_1 {\bd k}_1\otimes{\bd k}_1 +
 \alpha_2 {\bd k}_2\otimes{\bd k}_2 +
  \alpha_3 {\bd k}_3\otimes{\bd k}_3 ,
  \eeq
where $\alpha_j >0$ and $\{ {\bd k}_1,{\bd k}_2,{\bd k}_3\}$
is an orthonormal triad of vectors.
The six eigenvalues of ${\bd A}(0)$ therefore split into three pairs with opposite signs:
\begin{subequations}
\begin{align}
\lambda_1 &= - \sqrt{\alpha_1} = - \lambda_2,
\\
\lambda_3 &= - \sqrt{\alpha_2} = - \lambda_4,
\\
\lambda_5 &= - \sqrt{\alpha_3} = - \lambda_6.
\end{align}
\end{subequations}
The corresponding
  right and left  eigenvectors of ${\bd A}(0)$ are, respectively,
%\end{document}
\begin{align}\label{270}
 & {\bd r}_{2j-1}  =   \begin{pmatrix}   - \lambda_{2j-1} {\bd k}_j \\    {\bd k}_j  \end{pmatrix},
   \qquad
{\bd r}_{2j}  =  \begin{pmatrix}  - \lambda_{2j} {\bd k}_j \\   {\bd k}_j    \end{pmatrix} , &
   \nonumber    \\    & &  j=1,2,3.    \\
 & {\bd l}_{2j-1}  = \frac12  \begin{pmatrix}  - \lambda_{2j-1}^{-1} {\bd k}_j \\    {\bd k}_j
   \end{pmatrix} ,    \qquad
{\bd l}_{2j}  = \frac12  \begin{pmatrix}  - \lambda_{2j}^{-1} {\bd k}_j \\    {\bd k}_j
   \end{pmatrix} , & \nonumber
 \end{align}
 Note that  ${\bd l}_{i} \cdot {\bd r}_{j} = \delta_{ij}$.

 \begin{rmk}
 The property   $\alpha_j > 0$ implies that all six eigenvalues $\lambda_k$,
 $k = 1, 2, \ldots , 6$ of ${\bd A}(0)$  are real.
We assume that all the right and left eigenvectors of ${\bd A}(0)$
form a full set of linearly independent eigenvectors.  The  system
is therefore hyperbolic at the origin.  This assumption can be
expressed in terms of restrictions on the strain energy (e.g.
rank-one convexity), but we will not discuss this issue here. We
would like to emphasize, however, that we do admit the possibility
of coincident eigenvalues, i.e. non-strict hyperbolicity.  Moreover,
the main objective of this paper is the analysis of the case when
pairs of coincident eigenvalues correspond  to (quasi-)shear waves.
We call such waves \emph{degenerate}.
 \end{rmk}

\section{WNGO asymptotics}\label{sec3}

In this section we are interested in deriving the simplest nonlinear
evolution equations for the amplitudes of weakly nonlinear elastic
plane waves. Using the method of weakly nonlinear geometric optics
(WNGO) we will first derive equations for single waves and then for
coupled pairs of waves. {No attempt is made here to review  WNGO,
which has been widely developed and applied in many areas of
mechanics and physics  since the seminal work by  
\citet{Lax57}.  We refer the interested read to the review of
\cite{Hunter95}. }

For single waves two cases will be discussed.  In  considering the  system \rf{19} we  first  choose an eigenvalue $\lambda_s ({\bd w})$  of the matrix ${\bd A}({\bd w})$ in \rf{19} such that
\beq{280}
\left. \nabla_{\bd w} \lambda_s ({\bd w}) \cdot {\bd r}_s
\right|_{{\bd w}=0} \ne 0 ,
\eeq
where ${\bd r}_s$ is the  eigenvector corresponding to $\lambda_s$. This assumption, called \emph{genuine nonlinearity}, at the zero constant state is typically satisfied for longitudinal or quasi-longitudinal
waves. Using the perturbation method we will derive a nonlinear evolution for such waves
in Sec.~\ref{3.1}. Next, in Sec.~\ref{3.2} we also present the simplified evolution equations for single waves which do not satisfy the assumption \rf{280}. This is typical for shear or quasi-shear elastic waves.
Finally, in Sec.~\ref{3.3} we will also derive the evolution equations for the amplitudes of coupled waves that have  coincident speeds.

\subsection{Evolution equations for single (quasi-)longitudinal waves}\label{3.1}

Let us consider an initial-value problem
 for a quasilinear hyperbolic system \rf{19}:
\beq{28} \left\{
\begin{array}{l}
\displaystyle \frac{\partial{\bd w}^\epsilon} {\partial t} +{\bd
A}({\bd w}^\epsilon)
\frac{\partial{\bd w}^\epsilon}{\partial x}= {\bd 0} , \\[2ex]
\bd {w}^{\epsilon}(0,x)=  \epsilon\,\bd {w}_{1}(0,x,x/\epsilon),
\end{array}
\right.
\eeq
 where $\epsilon$  is a small parameter.
Note that   the high frequency
('fast') variable $ \frac{x}{\epsilon}$ appears in the initial condition.
The initial data are presumed  to have  compact support.

We consider the single wave weakly nonlinear geometrical optics
solution to \rf{28}
\beq{29}
{\bd w}^\epsilon (t,x) = \epsilon \sigma_s(t,x,\eta) {\bd r}_s + O( \epsilon^2),
\eeq
with an unknown amplitude $\sigma_s$  and a new independent variable
\beq{30}
\eta = \epsilon^{-1}(x-\lambda_s t).
\eeq
It is  assumed that the eigenvalue
$\lambda_s$ is distinct from the others, and  its
eigenvector is ${\bd r}_s$.

{
\begin{rmk}
Instead of using a `fast' variable, we may equally well use a `slow'
variable $\epsilon x$ and work with a solution of the form $w =
\bar{w}(x - \lambda t; \epsilon x)$, which represents a traveling
wave modified by nonlinear effects over a large  length scale. Using
this alternative starting point we could obtain the same results as
WNGO.
\end{rmk}}

Introducing the ansatz \rf{29} into \rf{28}, using a Taylor
expansion of ${\bd A} ({\bd w}^\epsilon )$ around ${\bd 0}$, we then
apply
the method of multiple-scale asymptotics.  This relies on treating $\eta$  as a new independent variable. % while performing all di®erentiations.
We sequentially collect terms of like powers in $\epsilon $  and
equating these  terms to zero  one finds that the solvability condition applied at
the O$(\epsilon )$ level yields a  nonlinear evolution equation for the unknown amplitude $\sigma_s$ (see \citep{Domanski00} for the details):
\beq{31}
\frac{\partial \sigma_s} {\partial t} +
\lambda_s \frac{\partial \sigma_s} {\partial x} +
\frac12 \Gamma_s \frac{\partial \sigma_s^2} {\partial \eta} = 0,
\eeq
where the coefficient of nonlinearity is
\beq{32}
\Gamma_s  = {\bd l}_s\cdot \big(
\nabla_{{\bd w}} {\bd A}({\bd w}) {\bd r}_s{\bd r}_s
\big)\big|_{{\bd w} = {\bd 0}}.
\eeq
One can show that $\Gamma_s $ is equal to the left hand side of \rf{280}, and hence $\Gamma_s \ne 0$.  An explicit equation for the nonlinearity coefficient
follows from eqs. \rf{14} and \rf{19} as
\begin{align}\label{51}
\Gamma_s  &= \frac12 \lambda_s^{-1}
{\bd k}_{\lfloor\! \frac{s+1}2\!\rfloor}
\cdot {\bd \Psi}\,
{\bd k}_{\lfloor\! \frac{s+1}2\!\rfloor}
{\bd k}_{\lfloor\! \frac{s+1}2\!\rfloor}
\nonumber \\
%&= \frac12 \lambda_s^{-1}\big( c_{abcdef}n_bn_dn_f k_a^sk_c^sk_e^s + 3 c_{abcd}n_bn_cn_d k_a^s\big) \nonumber \\
&= \frac12 \lambda_s^{-1}
c_{abcdef}n_bn_dn_f
k_a^{\lfloor\! \frac{s+1}2\!\rfloor}
k_c^{\lfloor\! \frac{s+1}2\!\rfloor}
k_e^{\lfloor\! \frac{s+1}2\!\rfloor}
 + \frac32 \lambda_s {\bd n}\cdot
{\bd k}_{\lfloor\! \frac{s+1}2\!\rfloor}
\, ,
\end{align}
where ${\lfloor ~ \rfloor}$ denotes the floor function, that is, the largest integer less than or equal to a given number. We also use
 the notation
${\bd k}_j \equiv   k_1^j {\bd e}_1 + k_2^j{\bd e}_2 + k_3^j{\bd e}_3$ in the Cartesian basis.
The parameter $\Gamma_s $ is well known in nonlinear elastodynamics - for instance, it governs the growth of elastic acceleration waves \citep{Norris91}.

\begin{examp}
Consider an isotropic elastic medium for which the strain energy is given by
\beq{Murn}
W = \frac12 (\lambda_L + 2 \mu_L) I_{\tiny \bd E}^2 -  2 \mu_L  I\!I_{\tiny \bd E}
+ \frac13 (l_M + 2 m_M) I_{\tiny \bd E}^3
-  2 m_M  I_{\tiny \bd E} I\!I_{\tiny \bd E}  + n_M I\!I\!I_{\tiny \bd E} ,
\eeq
where $\lambda_L$, $ \mu_L$ are second order Lam\'e constants, $l_M$, $m_M$
and $n_M$ are third order Murnaghan constants\footnote{Relations between the three Murnaghan constants and  alternative triads of third order constants for isotropic solids are listed by \cite{norris1993a}.}, and the strain invariants are
\beq{inv}
I_{\tiny \bd E}= \tr {\bd E} ,
\qquad
 {I\!}I_{\tiny \bd E}  = \frac{1}{2}\left[(\tr {\bd E})^2 - \tr {\bd E}^2\right],
 \qquad
  I\!I\!I_{\tiny \bd E}  = \det {\bd E}.
\eeq In this case the two coefficients appearing in the evolution
equation \rf{31} are \beq{coef1} \lambda_s = - \sqrt{\lambda_L + 2
\mu_L} \quad \text{and} \quad \Gamma_s = \frac{3(\lambda_L + 2
\mu_L) + 2(l_M + 2 m_M) } {2 \sqrt{\lambda_L + 2 \mu_L}}. \eeq
Hence, we see that in the isotropic elastic medium the coefficients
in the evolution equation for  longitudinal waves are determined by
the two second order and the two third order constants, $\lambda_L$,
$\mu_L$ and $l_M$, $ m_M$, respectively.
\end{examp}

\subsection{Evolution equations for single (quasi-)shear waves}\label{3.2}

 We now present the simplest nonlinear evolution equation for the amplitude $\sigma_s$
of a single weakly nonlinear (quasi-)shear elastic wave for which $\Gamma_s = 0$.
The case when $\Gamma_s = 0$
requires a different scaling from that of \rf{30},
one appropriate to cubic nonlinearity as the leading term.
The procedure is described in detail by \citet{Domanski00}.
The WNGO solution for the single wave has the same formal expansion as in eq. \rf{29},
but now, crucially,  $\eta = \epsilon^{-2}(x-\lambda_s t)$.
The modified asymptotics leads to the following governing equation
for $\sigma_s$:
\beq{34}
\frac{\partial \sigma_s} {\partial t} +
\lambda_s \frac{\partial \sigma_s} {\partial x} +
\frac13 G_s \frac{\partial \sigma_s^3} {\partial \eta} = 0,
\eeq
with nonlinearity coefficient
\beq{35}
G_s  = \frac12 \lambda_s^{-1}
{\bd k}_{\lfloor\! \frac{s+1}2\!\rfloor}
\cdot \big( 3{\bd \Psi}\,
{\bd k}_{\lfloor\! \frac{s+1}2\!\rfloor}
{\bd q} + {\bd \Pi}\,
{\bd k}_{\lfloor\! \frac{s+1}2\!\rfloor}
{\bd k}_{\lfloor\! \frac{s+1}2\!\rfloor}
{\bd k}_{\lfloor\! \frac{s+1}2\!\rfloor}
\big),
\eeq
where the vector  ${\bd q}$ is orthogonal to ${\bd k}_s$ and satisfies
\beq{36}
( {\bd \Lambda}- \lambda_s^2{\bd I}) {\bd q} +
{\bd \Psi}\,
{\bd k}_{\lfloor\! \frac{s+1}2\!\rfloor}
{\bd k}_{\lfloor\! \frac{s+1}2\!\rfloor}
 ={\bd 0}.
\eeq

\begin{examp}
For an isotropic elastic medium with strain energy  given by
\rf{Murn} the two coefficients  in the evolution equation \rf{34}
are \beq{coef2} \lambda_s = - \sqrt{\mu_L} \quad \text{and} \quad
G_s = \frac{3}{4 \lambda_s} \bigg[ \frac{\lambda_L \, \mu_L +  (
\mu_L+ m_M)^2 }{\lambda_L +  \mu_L} \bigg]. \eeq Hence, in the
isotropic case, the coefficients in the evolution equation for shear
waves are determined by the two second order and only one of the
third order constants, $\lambda_L$, $\mu_L$, and $ m_M$,
respectively.
\end{examp}

\begin{rmk}
{
By applying the method of characteristics
we reduce the differentiation $\frac{\partial}{\partial t}
+ \lambda  \frac{\partial}{\partial x}$
to the  differentiation $\frac{\partial}{\partial \tau}$ %along the characteristic
where $\tau = t - \lambda x$
is a characteristic variable.  In this way the partial differential equation
\beq{37}
 \frac{\partial \sigma}{\partial t} + \lambda \frac{\partial
\sigma}{\partial x} + \frac{1}{j}\Gamma\frac{\partial
\sigma^j}{\partial\eta} = 0 ,
\end{equation}
for  the function of three independent variables
%constants $\lambada$ and $\Gamma$ and the function of three independent variables
$\sigma = \sigma(t, x, \eta)$  transforms
to the partial differential equation
\beq{38}
  \frac{\partial \tilde{\sigma}}{\partial \tau}
+    \frac{1}{j}\Gamma\frac{\partial
\tilde{\sigma}^j}{\partial \eta} = 0 ,
\end{equation}
for the function of two independent variables
$ \tilde{\sigma} = \tilde{\sigma}(\tau, \eta)$.
Here  $j = 1,2,3,...$ is a natural number. }
When $j = 2$, as occurs for the longitudinal wave evolution equation \rf{31},  we call the equation \rf{38}
 the \emph{inviscid Burgers equation}.  Similarly,
 $j = 3$ for the (quasi-) shear wave in  \rf{34}, the  equation \rf{38} is called
 the \emph{modified inviscid Burgers equation}.

\end{rmk}

\subsection{Degenerate plane waves: acoustic axes}\label{3.3}
In this section we derive the simplest nonlinear evolution equations for the amplitudes of a pair of weakly nonlinear elastic waves in the case when these waves have coincident wave speeds.

Consider the Christoffel tensor ${\bd \Lambda}$ from \rf{231}.\\
{\bf Definition}: We say that the eigenvalues of ${\bd \Lambda}$ are
{\it degenerate}, if \beq{40} {\bd \Lambda} =
 \alpha_1 ( {\bd I} - {\bd k}_3\otimes{\bd k}_3 )
 + \alpha_3 {\bd k}_3\otimes{\bd k}_3 ,
  \eeq
  that is, if $ \alpha_1= \alpha_2$ in \rf{231}.
In such a situation we say that  $\bd n$ is an {\it acoustic axis}
 (see \citet{khatkevich1962} or \citet{Norris04b}).

Let $\{ {\bd k}_1,\, {\bd k}_2,\, {\bd k}_3\}$  be an orthonormal
triad of vectors and define the left and right
eigenvectors of ${\bd A} ({\bd 0})$ as before, see eq. \rf{270}. Then
${\bd A}({\bd 0})$ has two pairs of coincident eigenvalues:
$\lambda_s =\lambda_{s+2}$ for $s=1$ and $s=2$.   We now consider
the following ansatz for the initial value problem \rf{28}: \beq{41}
{\bd w}(t,x) = \epsilon \bigg( \sigma_s(t,x,\eta) {\bd r}_s
+\sigma_{s+2}(t,x,\eta) {\bd r}_{s+2}\bigg) +O( \epsilon^2), \qquad
s=1,2, \eeq with $\eta = \epsilon^{-1}(x-\lambda{_s} t)$,
%{$\lambda \equiv \lambda_s = \lambda_{s+2}$}
and where $\sigma_s$ and $\sigma_{s+2}$ are the unknown amplitudes.
Inserting \rf{41} into \rf{28} and applying the multiple scale
asymptotic methods described in Section \ref{3.1} (see
\citet{Domanski00}) we obtain the following pair of coupled
evolution equations for the amplitudes $\sigma_s$ and
$\sigma_{s+2}$:
\begin{subequations}\label{04}
\bal{04a} \frac{\partial \sigma_s} {\partial t} + \lambda{_s}
\frac{\partial \sigma_s} {\partial x} + \frac12 \bigg(
\Gamma^{s}_{s,s} \frac{\partial \sigma_s^2} {\partial \eta} +2
\Gamma^{s}_{s,s+2} \frac{\partial (\sigma_s\sigma_{s+2})} {\partial
\eta} +\Gamma^{s}_{s+2,s+2} \frac{\partial \sigma_{s+2}^2} {\partial
\eta} \bigg) &= 0,
\\
\frac{\partial \sigma_{s+2}} {\partial t} + \lambda{_s}
\frac{\partial \sigma_{s+2}} {\partial x} + \frac12 \bigg(
\Gamma^{s+2}_{s,s} \frac{\partial \sigma_s^2} {\partial \eta} +2
\Gamma^{s+2}_{s,s+2} \frac{\partial (\sigma_s\sigma_{s+2})}
{\partial \eta} +\Gamma^{s+2}_{s+2,s+2} \frac{\partial
\sigma_{s+2}^2} {\partial \eta} \bigg) &= 0,\label{04b}
\end{align}
\end{subequations}
where the interaction coefficients are, in general,
\beq{44}
\Gamma^j_{p,q}  = {\bd l}_j\cdot \big(
\nabla_{{\bd w}} {\bd A}({\bd w}) {\bd r}_p{\bd r}_q
\big)|_{{\bd w} = {\bd 0}}. %= \frac12 \lambda_j^{-1}{\bd k}_j\cdot {\bd \Psi}{\bd k}_p{\bd k}_q\, .
\eeq The assumption of hyperelasticity (\ref{02}) implies that these
coefficients have the following symmetry property
\beq{+20}\Gamma^j_{p,q} = \Gamma^j_{q,p}.\eeq
{Moreover the formulas
\rf{270} imply that $\Gamma^j_{p,q} = -\Gamma^{j + 1}_{p,q}$ for
j = 1,3,5.}
In our case we have \beq{441} \Gamma^j_{p,q}  =  \frac12
\lambda_j^{-1}{\bd k}_{\lfloor\! \frac{j+1}2\!\rfloor} \cdot {\bd
\Psi}\, {\bd k}_{\lfloor\! \frac{p+1}2\!\rfloor}{\bd k}_{\lfloor\!
\frac{q+1}2\!\rfloor}\, . \eeq
Using Cartesian components  we can
express the interaction coefficients as follows
\begin{align}
\Gamma^j_{p,q} &=  \frac12 \lambda_j^{-1}\big[
c_{abcdef}n_bn_dn_f
k_a^{\lfloor\! \frac{j+1}2\!\rfloor}
k_c^{\lfloor\! \frac{p+1}2\!\rfloor}
k_e^{\lfloor\! \frac{q+1}2\!\rfloor}
\nonumber \\ & \qquad  \qquad  \qquad
+  c_{abcd}n_bn_cn_d
\big( k_a^{\lfloor\! \frac{j+1}2\!\rfloor} \delta_{pq}
+ k_a^{\lfloor\! \frac{p+1}2\!\rfloor}\delta_{jq}
+ k_a^{\lfloor\! \frac{q+1}2\!\rfloor}\delta_{jp}\big)
\big]
\nonumber \\
&= \frac12 \lambda{_j}^{-1} c_{abcdef}n_bn_dn_f k_a^{\lfloor\!
\frac{j+1}2\!\rfloor} k_c^{\lfloor\! \frac{p+1}2\!\rfloor}
k_e^{\lfloor\! \frac{q+1}2\!\rfloor} \nonumber \\ & \qquad  \qquad
\qquad +  \frac12\lambda{_j} {\bd n}\cdot \big(
    {\bd k}_{\lfloor\! \frac{j+1}2\!\rfloor} \delta_{pq}
+   {\bd k}_{\lfloor\! \frac{p+1}2\!\rfloor} \delta_{jq}
+   {\bd k}_{\lfloor\! \frac{q+1}2\!\rfloor} \delta_{jp}\big)
\, .
\end{align}
%{where j,p,q can take values 1,2,3,4.}
 Therefore the coefficients have in our case, in addition to the general
 property
\rf{+20}, the following indicial symmetries: \beq{+2} \Gamma^j_{p,q}
= \Gamma^p_{j,q} . \eeq This makes them \emph{totally symmetric}
under the interchange of indices. In particular, note that
\begin{subequations}\label{54}
\bal{54a}
\Gamma^{s+2}_{s,s} = \Gamma^{s}_{s,s+2} =
  \frac12 \lambda{_s}^{-1}
c_{abcdef}n_bn_dn_f k_a^{\lfloor\! \frac{s+1}2\!\rfloor}
k_c^{\lfloor\! \frac{s+1}2\!\rfloor} k_e^{\lfloor\!
\frac{s+3}2\!\rfloor} +  \frac12\lambda{_s}  {\bd n}\cdot{\bd
k}_{\lfloor\! \frac{s+3}2\!\rfloor} & \equiv \Gamma^{s+2}_{s},
\\
\Gamma^{s}_{s+2,s+2} = \Gamma^{s+2}_{s,s+2} =
  \frac12 \lambda{_s}^{-1}
c_{abcdef}n_bn_dn_f k_a^{\lfloor\! \frac{s+1}2\!\rfloor}
k_c^{\lfloor\! \frac{s+3}2\!\rfloor} k_e^{\lfloor\!
\frac{s+3}2\!\rfloor} +  \frac12\lambda{_s}  {\bd n}\cdot{\bd
k}_{\lfloor\! \frac{s+1}2\!\rfloor} &\equiv \Gamma^{s}_{s+2}.
\label{54b}
\end{align}
\end{subequations}

In summary:
\begin{lem}\label{lem1}
The evolution equations  for the  amplitudes of  shear waves propagating along an acoustic axis are
\begin{subequations}\label{-2}
\bal{-2a} \frac{\partial \sigma_s} {\partial t} + \lambda{_s}
\frac{\partial \sigma_s} {\partial x} + \frac12 \bigg( \Gamma_s
\frac{\partial \sigma_s^2} {\partial \eta} +2 \Gamma^{s+2}_{s}
\frac{\partial (\sigma_s\sigma_{s+2})} {\partial \eta}
+\Gamma^{s}_{s+2} \frac{\partial \sigma_{s+2}^2} {\partial \eta}
\bigg) &= 0,
\\
\frac{\partial \sigma_{s+2}} {\partial t} + \lambda{_s}
\frac{\partial \sigma_{s+2}} {\partial x} + \frac12 \bigg(
\Gamma^{s+2}_{s} \frac{\partial \sigma_s^2} {\partial \eta} +2
\Gamma^{s}_{s+2} \frac{\partial (\sigma_s\sigma_{s+2})} {\partial
\eta} +\Gamma_{s+2} \frac{\partial \sigma_{s+2}^2} {\partial \eta}
\bigg) &= 0.\label{-2b}
\end{align}
\end{subequations}
The  nonlinear terms in the equations involve four  coefficients:
$\Gamma_s$ and $\Gamma_{s+2}$ from eq. \rf{51},  $\Gamma^{s+2}_{s}$ and
$\Gamma_{s+2}^{s}$ from eqs. \rf{54}, the latter
  two of which determine the coupling between the equations.
\end{lem}

\section{Simplification along symmetry axes}\label{sec4}

The previous  analysis shows that there are at most four independent
coefficients appearing in  the nonlinear terms  of the shear elastic
wave's coupled system \rf{-2}. We will demonstrate  the following:
\begin{lem}\label{lem2}
The number $r$ of  coefficients in  the nonlinear terms in the
coupled equations {\rf{-2}} for a pair of shear waves is always
reduced if the direction of propagation  $\bd n$  is a symmetry
{acoustic} axis. Specifically,
\begin{verse}
 $r=2$ for propagation along a 2-fold symmetry acoustic axis: {$\Gamma_s=\Gamma^s_{s+2}=0$;}  \\
 $r=1$  for a 3-fold symmetry acoustic axis: {$\Gamma_s=\Gamma^s_{s+2}=0$} and {$\Gamma_{s+2} +\Gamma_s^{s+2}=0$};  \\
 $r=0$  for an acoustic axis of 4-fold or higher symmetry: {$\Gamma_s=\Gamma^s_{s+2}=\Gamma_{s+2} =
 \Gamma_s^{s+2}=0$.}
 %{$\Gamma_{s+2} =0$}.
 \end{verse}
\end{lem}%{tabble}

\subsection{Two-fold axis}
We say that the propagation direction is a \emph{two-fold axis of symmetry} if it lies in a plane of symmetry of a monoclinic solid.  Let ${\bd e}$ be the normal to the plane of monoclinic symmetry, then ${\bd e}$ also belongs to the plane of degenerate wave vectors.  Let ${\bd e}$ coincide  with ${\bd e}_1$, and be parallel to ${\bd k}_{\lfloor\! \frac{s+1}2\!\rfloor}$.
It then   follows that ${\bd n}\cdot {\bd k}_{\lfloor\! \frac{s+1}2\!\rfloor}=0$, and from eqs. \rf{51} and
\rf{54b}, we have
\begin{subequations}\label{56}
\bal{56a} \Gamma^s_{ s}&\equiv \Gamma^s_{s,s} = \frac12
\lambda{_s}^{-1} c_{abcdef}n_bn_dn_f k^{\lfloor\!
\frac{s+1}2\!\rfloor}_a k^{\lfloor\! \frac{s+1}2\!\rfloor}_c
k^{\lfloor\! \frac{s+1}2\!\rfloor}_e,
\\
\Gamma^s_{s+2}&\equiv  \Gamma^s_{s+2,s+2}= \frac12
\lambda{_s}^{-1} c_{abcdef}n_bn_dn_f k_ck_e k^{\lfloor\!
\frac{s+1}2\!\rfloor}_a k^{\lfloor\! \frac{s+3}2\!\rfloor}_c
k^{\lfloor\! \frac{s+3}2\!\rfloor}_e .
\end{align}
\end{subequations}
Each term in these expressions involves an element of $c_{abcdef}$
with the index $1$ occurring either once or thrice.  But by
definition of a plane of symmetry these elements vanish, and hence
the two elements $\Gamma_s$ and $\Gamma^s_{s+2}$ vanish.  The
canonical form of the evolution equations is
\begin{subequations}\label{82}
\bal{82a} \frac{\partial \sigma_s} {\partial t} + \lambda{_s}
\frac{\partial \sigma_s} {\partial x} +
 \Gamma^{s+2}_{s} \frac{\partial (\sigma_s\sigma_{s+2})} {\partial \eta}
&= 0,
\\
\frac{\partial \sigma_{s+2}} {\partial t} + \lambda{_s}
\frac{\partial \sigma_{s+2}} {\partial x} + \frac12 \bigg(
\Gamma^{s+2}_{s} \frac{\partial \sigma_s^2} {\partial \eta}
+\Gamma_{s+2} \frac{\partial \sigma_{s+2}^2} {\partial \eta} \bigg)
&= 0.\label{82b}
\end{align}
\end{subequations}

\subsection{Three-fold axis}
 The generic configuration in this case is propagation along the axis of a trigonal material.
 This possesses three planes of symmetry each containing the axis of symmetry, and mutually
 at $120^\circ$ to one another. Reasoning similarly to the case of the two-fold symmetry axis,
 we get that {$\Gamma_s =\Gamma^s_{s+2} = 0 $}, but here we have moreover that
 \beq{00}
 \Gamma_{s+2} +\Gamma_s^{s+2}=0 ,
 \eeq
 see below for  details. Therefore the canonical form of the evolution equations for a three-fold symmetry axis is
 \begin{subequations}\label{83}
\bal{83a} \frac{\partial \sigma_s} {\partial t} + \lambda{_s}
\frac{\partial \sigma_s} {\partial x} -
 \Gamma_{s+2} \frac{\partial (\sigma_s\sigma_{s+2})} {\partial \eta}
&= 0,
\\
\frac{\partial \sigma_{s+2}} {\partial t} + \lambda{_s}
\frac{\partial \sigma_{s+2}} {\partial x} - \frac12 \Gamma_{s+2}
\bigg( \frac{\partial \sigma_s^2} {\partial \eta} -\frac{\partial
\sigma_{s+2}^2} {\partial \eta} \bigg) &= 0.\label{83b}
\end{align}
\end{subequations}

Hence only one coefficient $\Gamma_{s+2}$ characterizes the nonlinear terms in the evolution equations for a pair of degenerate plane waves in the case when the propagation direction is a three-fold symmetry axis.

\begin{rmk}
The system \rf{83}  can be transformed into the single complex Burgers
equation. This equation was studied in the Noelle's Ph.D. thesis
\citep{Noelle90}.  It was shown there that the solutions of the complex
Burgers equation become singular in finite time for a large class of
initial data. The shocks which develop are of nonclassical type,  that
is, they do not satisfy Lax's conditions. This is due to the fact
that too few characteristics enter the shock front in comparison to
compressive shocks which satisfy Lax's condition.
\end{rmk}

\subsubsection{Proof of equation \rf{00}}

The basis for the proof rests on the property \rf{+2} combined with
the three-fold symmetry. Using the same arguments as for the
two-fold axis, the three-fold symmetry is associated with three
planes of monoclinic symmetry with normals all perpendicular to
${\bd n}$ and since each normal lies in the plane of degenerate wave
vectors, it follows that the degenerate wave vectors are orthogonal
to the axis and are therefore pure transverse waves.

In order to simplify matters, recall that for propagation along an acoustic axis $s=1$ or $s=2$ (see eq. \rf{41}), and therefore the coefficients
$\Gamma^j_{p,q}$ may be identified with the elements of a
 totally symmetric third order tensor of dimension 2.
 Thus, {
\beq{22}
 {\bd g} =  g_{\alpha \beta \gamma}
 {\bd k}_\alpha \otimes{\bd k}_\beta \otimes{\bd k}_\gamma ,
\eeq } where  lower case Greek subscripts take values 1 and 2,
{and, for instance,} {\beq{-9} g_{\alpha \beta \gamma} =
c_{abcdef}n_bn_dn_f k_a^\alpha k_c^\beta k_e^\gamma
 +  c_{abcd}n_bn_cn_d
\big( k_a^{\alpha} \delta_{\beta \gamma} + k_a^{\beta}\delta_{\gamma
\alpha} + k_a^{\gamma}\delta_{\alpha \beta}\big)
 \eeq}
%{and, for instance,
% \beq{-9}
%g_{\alpha \beta \gamma} =  c_{abcdef}n_bn_dn_f
%k_a^\alpha
%k_c^\beta
%k_e^\gamma
% +   \lambda^2 {\bd n}\cdot \big(
%    {\bd k}_\alpha \delta_{\beta \gamma}
%+   {\bd k}_\beta \delta_{\gamma\alpha}
%+   {\bd k}_\gamma \delta_{\alpha\beta}\big).
%\eeq}
 {In this case the relationship with  the nonlinearity coefficients $\Gamma^j_{p,q}$ is defined by
 \beq{-3}
 \Gamma^j_{p,q} \equiv  {\frac{1} {2 \lambda_j}} g_{\lfloor\! \frac{j+1}2\!\rfloor
 \lfloor\! \frac{p+1}2\!\rfloor \lfloor\! \frac{q+1}2\!\rfloor  },
 \eeq}{where j,p,q take values 1,2,3,4.}
%Also,   ${\bd e}_1,  {\bd e}_2$ are, for the moment, an arbitrary pair of orthonormal basis vectors orthogonal to the axis.
The totally symmetric property means that $g_{\alpha \beta \gamma}$ is
unchanged under any permutation of the three indices, and hence
there are at most four independent elements. Consider the change of
basis {
\beq{23} {\bd k}_1'= \cos\theta {\bd k}_1 + \sin\theta {\bd
k}_2, \qquad {\bd k}_2'= -\sin\theta {\bd k}_1 + \cos\theta {\bd
k}_2, \eeq and define \beq{302} g_{\alpha \beta \gamma}(\theta) =
{\bd g} {\bd k}_\alpha' {\bd k}_\beta'{\bd k}_\gamma',
\eeq }
then
\beq{301}
\begin{pmatrix}
g_{111}(\theta)\\
g_{222}(\theta)\\
g_{112}(\theta)\\
g_{122}(\theta)
\end{pmatrix}
=
\begin{bmatrix}
c^3 & s^3 & 3c^2 s & 3c s^2 \\
-s^3 & c^3 & 3cs^2  & -3c^2s \\
-c^2s & cs^2 & c^3-2cs^2 & 2c^2s-s^3\\
cs^2 & c^2s & s^3-2c^2s & c^3-2cs^2
\end{bmatrix}
\begin{pmatrix}
g_{111}(0)\\
g_{222}(0)\\
g_{112}(0)\\
g_{122}(0)
\end{pmatrix},
\eeq
where $c= \cos\theta$, $s= \sin\theta$.   We note the property $g_{\alpha \beta \gamma}(\theta) = - g_{\alpha \beta \gamma}(\theta +\pi)$ for  each element  $g_{\alpha \beta \gamma}(\theta)$.

 Let ${\bd e}_1$ be the normal to one of the three planes, then the coefficients
$g_{111}(0)$ and $g_{122}(0)$ vanish. Similarly, $g_{111}(\pm
\frac23 \pi) $ and $g_{122}(\pm \frac23 \pi)$ must vanish. Using
\rf{301} with $g_{111}(0)=g_{122}(0)=0$, we find that \beq{66}
g_{111}(\pm \tfrac23 \pi) = 3g_{122}(\pm \tfrac23 \pi) = \pm
\frac{3\sqrt{3}}{8} \big( g_{112}(0)+g_{222}(0)\big), \eeq and hence
the 3-fold symmetry requires that \beq{67} g_{112}(0)+g_{222}(0)= 0.
\eeq This is precisely eq. \rf{00}.   In summary, the evolution
equations remain coupled, but are characterized by a single
nonlinearity parameter {and take the form \rf{83}}.

\subsection{Four-fold axis}
If the axis is one of 4-fold symmetry, then the same arguments as above imply that the degenerate wave-vectors are orthogonal to the axis, and are thus pure transverse waves.  The same arguments also imply that the elements of $\bd \Gamma$ associated with the second plane of symmetry must also vanish, and hence all four are zero.  Thus,  ${\bd \Gamma} = 0 $ and there is no coupling.

The same reasoning applies to axes of higher symmetry, since all such axes are equivalent to an axis of transverse isotropy.
Therefore the canonical form of the evolution equations for a symmetry  axis with
four-fold or higher symmetry is
 \begin{subequations}\label{84}
\bal{84a} \frac{\partial \sigma_s} {\partial t} + \lambda{_s}
\frac{\partial \sigma_s} {\partial x} &= 0,
\\
\frac{\partial \sigma_{s+2}} {\partial t} + \lambda{_s}
\frac{\partial \sigma_{s+2}} {\partial x} &= 0.\label{84b}
\end{align}
\end{subequations}

\subsection{Applications to cubic crystals}

Let us consider a cubic crystal of class $m3m$ in which the strain
energy $W$ is defined by three second order and six third order
elastic constants \citep{Domanski00,Domanski00b,Domanski08b}, (see
also the Appendix): \beq{61} W = W (c_{11},\,c_{12},\,c_{44},\,
    c_{111},\, c_{112},\,c_{144},\,c_{123},\,c_{166},\,c_{456}).
    \eeq
    We focus on three particular directions of plane waves'
    propagation:  $[1\, 0\, 0],  [1\, 1\, 0]$ and  $[1\, 1\, 1]$.

%We derive explicit formulas for the evolution equations for quasi-longitudinal and quasi-transverse in both decoupled and coupled cases.
\begin{examp}\label{ex3} Consider first the case of propagation
direction ${\bd n} = [1\,0\,0]$.  This direction is along a
four-fold symmetry acoustic axis.  The shear wave speeds are given
by \beq{62a} \lambda_1 = \lambda_3 = - \sqrt{ c_{44} } = -\lambda_2
= -\lambda_4. \eeq Hence, the eigenvalues $\lambda_1$, $\lambda_2$,
$\lambda_3$ and $\lambda_4$ are coupled but the second order
nonlinearity coefficient $\Gamma_s$ vanishes for $s=1,2,3,4$.
Therefore there is no quadratically nonlinear coupling in the
evolution equations for shear waves. One can show that propagation
of the shear plane waves is  described  by the evolution equation
with cubic  nonlinearity, eq. \rf{34} \citep{Domanski00}. {The
coefficients at cubic nonlinearity in this equation are as follows:
\beq{62b} G_s = - \displaystyle\frac{3}{4 \lambda_s } \left [ \frac{
2 (c_{11} + c_{166})^2 + c_{44} ( c_{11} + c_{166})  + c_{166}
(c_{44} + c_{166})}{ c_{11} - c_{44}} \right], \,\,\,\,\,\, s =
1,2,3,4. \eeq }
\end{examp}

\begin{examp}\label{ex4}
Consider now the case of propagation direction ${\bd n} =
\frac1{\sqrt{2}} [1\,1\,0]$.  This direction is along a two-fold
symmetry axis and is  {\it not} an acoustic axis.  The shear wave
speeds are given by \beq{62} \lambda_1 = - \sqrt{\frac{ c_{11}
-c_{12} }{2} } = -\lambda_2, \qquad \lambda_3 = - \sqrt{c_{44 }} =
-\lambda_4. \eeq Hence, the eigenvalues $\lambda_1$, $\lambda_2$,
$\lambda_3$ and $\lambda_4$ are distinct and the second order
nonlinearity coefficient $\Gamma_s$ vanishes for $s=1,2,3,4$. There
is no quadratic coupling between the shear waves, and furthermore
one can show that propagation of the shear plane waves is  described
by the evolution equation with cubic  nonlinearity, eq. \rf{34}
\citep{Domanski00}. { The coefficients in this equation are:
\beq{62c} G_s = - \displaystyle \frac{3}{16 \lambda_s } \left [
\frac{ [c_{11} - c_{12} + \frac{1}{2} (c_{111} - c_{112})]^2 + 2
(c_{11} - c_{12})(c_{12} + c_{44})}{ c_{12} + c_{44}} \right ],
\,\,\,\,\,\, s = 1,2; \eeq \\
 \beq{62d} G_s = - \displaystyle
\frac{3}{16 \lambda_s } \left [ \frac{ (c_{11} - c_{12} + c_{144} +
c_{166} + 2 c_{456})^2 + 2 (c_{11} - c_{12})(c_{12} + c_{44})}{
c_{12} + c_{44}} \right ], \,\,\,\,\,\, s = 3,4. \eeq }
\end{examp}

\begin{examp}\label{ex5}
Finally let us consider the case of propagation along an axis of
three-fold symmetry, ${\bd n} = \frac1{\sqrt{3}} [1\,1\,1]$, which
is also an acoustic axis.  In this case the shear wave amplitudes
are coupled and are described by the coupled evolution equations
\rf{83} with (see the Appendix)
\begin{subequations}\label{63}
\bal{63a} &\lambda_1 = \lambda_3 = - \sqrt{\frac{ c_{11}
-c_{12}+c_{44} }{3} } = -\lambda_2 = -\lambda_4,
\\
 & \Gamma_s =  \Gamma_1 = 0 = \Gamma_2,
 \\
 & \Gamma_{s + 2} = \Gamma_3 = \frac1{18\sqrt{2}\lambda_1 }
 \big[
 c_{111} +2c_{123}-2 c_{456}  -3(c_{112}-c_{144}+c_{166 })
 \big]= - \Gamma_4 .
\end{align}
\end{subequations}

\end{examp}

\section{Degenerate transverse waves in the absence of symmetry}\label{sec5}

\subsection{The principal result}

In this Section we consider the general case of {quasi-transverse degenerate wave vectors}.  No symmetry is assumed.  We examine the possibility that the coupled nonlinear wave equations for the two amplitudes decouple, and derive a general condition that is both necessary and sufficient for this to occur.

The main result is the following:
\begin{lem}\label{lemx}
 A pair of {quasi-transverse}  waves propagating along an acoustic axis are decoupled if and only if the nonlinearity coefficients satisfy the  identity
\beq{-6}
\Gamma_s\Gamma^s_{s+2}+\Gamma_{s+2}\Gamma_s^{s+2}
-(\Gamma^s_{s+2})^2- (\Gamma_s^{s+2})^2
=0.
\eeq
 If this condition is met then there is a coordinate transformation for which the coupling terms disappear and the degenerate transverse waves satisfy separate but different evolution equations:
\begin{subequations}\label{-5}
\bal{-5a} \frac{\partial \sigma_s} {\partial t} + \lambda{_s}
\frac{\partial \sigma_s} {\partial x} +\frac12 \Gamma_s'
\frac{\partial \sigma_s^2 } {\partial \eta}  &=0,
  \\
\frac{\partial \sigma_{s+2}} {\partial t} + \lambda{_s}
\frac{\partial \sigma_{s+2}} {\partial x} +\frac12 \Gamma_{s+2}'
\frac{\partial \sigma_{s+2}^2} {\partial \eta}  &=0.
\end{align}
\end{subequations}
\end{lem}

{
\begin{rmk}
The condition \rf{-6} can be expressed in terms of the 2-dimensional
third order symmetric tensor $\bd g$ as (see eq. \rf{271}) \beq{+6}
g_{\alpha \beta\gamma}g_{\alpha \beta\gamma}- g_{\alpha
\beta\beta}g_{\alpha \gamma\gamma}=0 . \eeq Based upon eq. \rf{-9}
this may be interpreted as a specific constraint on  the third order
moduli  involving the direction $\bd n$ of the acoustic axis.  It
also depends upon the second order moduli through the {common}
eigenvalue, but this dependence disappears when the
degenerate waves are purely transverse, in which case \rf{-6} is
strictly a relation between the third order moduli.
\end{rmk}}

\subsection{Derivation of Lemma \ref{lemx}}

{We use} the property seen previously in Section \rf{sec4} that the
coefficients $\Gamma^j_{p,q}$ are the elements of a third order
totally symmetric tensor in two dimensions, ${\bd g}$ of eq.
\rf{22}. { Let $\lambda_1 = \lambda_2 = \lambda$.} The coupled wave
equations can be expressed in terms of a 2-vector for the
displacement gradient vector of eq. \rf{153}: $ {\bd m}  =
\sigma_s{\bd k}_s+ \sigma_{s+2}{\bd k}_{s+2}$,  or ${\bd m}=m_1{\bd
e}_1+m_2{\bd e}_2$, where ${\bd e}_1,{\bd e}_2$ are orthonormal in
the plane of ${\bd k}_s$, ${\bd k}_{s+2}$:
\begin{align}
\frac{\partial m_1} {\partial t} + \lambda{_1} \frac{\partial
m_1} {\partial x} +\frac12 \frac{\partial } {\partial \eta} \big(
g_{111}m_1^2 +2g_{112}m_1m_2+ g_{122}m_2^2\big)&=0,
\nonumber \\
\frac{\partial m_2} {\partial t} + \lambda{_2} \frac{\partial
m_2} {\partial x} +\frac12 \frac{\partial } {\partial \eta}
\big(g_{112}m_1^2 + 2g_{122}m_1m_2 + g_{222}m_2^2\big)&=0, \nonumber
\end{align}
or
\beq{211}
\frac{\partial {\bd m}} {\partial t} +
\lambda \frac{\partial {\bd m}} {\partial x} +\frac12
\frac{\partial } {\partial \eta} \big( {\bd g}{\bd m}{\bd m}\big)={\bd 0}.
\eeq
The coupling between $m_1$ and $m_2$ vanishes iff the two elements $g_{112}$ and
$g_{122}$  are simultaneously zero.  Even if these are non-zero there might  exist a coordinate transformation in which the transformed quantities vanish.  It is this possibility that we seek.

Thus, we consider the possibility that there is some angle of rotation $\theta$ such that $g_{112}(\theta)=0$ and
$g_{122}(\theta)=0$ where
$g_{\alpha \beta \gamma}(\theta)$ are defined by \rf{302}.   These two conditions are simultaneously satisfied if the 2-vector $ {\bd g} {\bd e}_1' {\bd e}_2'$ vanishes. Under the rotation \rf{23} we have
\beq{24}
 {\bd g} {\bd e}_1' {\bd e}_2'= {\bd a}\cos 2\theta + {\bd b}\sin 2\theta,
 \eeq
where
\beq{25}
{\bd a} =  {\bd g} {\bd e}_1 {\bd e}_2,
\qquad
 {\bd b}=  \frac12( {\bd g} {\bd e}_2 {\bd e}_2 - {\bd g} {\bd e}_1 {\bd e}_1 ).
 \eeq
The form of \rf{24} indicates that the vector
 ${\bd g} {\bd e}_1' {\bd e}_2'$ can be zero if and only if
 ${\bd a}$ and ${\bd b}$ are parallel.
Thus, the general condition for no coupling is
\beq{26}
\mbox{no coupling}\quad \Leftrightarrow \quad
{\bd a}\times  {\bd b} = {\bd 0},
\eeq
which is equivalent to
\beq{27}
\mbox{no coupling}\quad \Leftrightarrow \quad
 \mu \equiv g_{112}^2+g_{122}^2-g_{112}g_{222}- g_{122}g_{111} = 0.
\eeq
This is precisely the result of eq. \rf{-6} in Lemma \ref{lemx}.
We will show next that the quantity $\mu$  of eq. \rf{27} is an invariant, independent of the coordinates used.

\subsection{Tensor properties of the nonlinearity coefficients}

The third order tensor $ {\bd g}$ is 2-dimensional and totally symmetric, that is, the Cartesian components are unchanged under permutation of the indices.
\citet{Jerphagnon70}  considered the general form of third order tensors in 3-dimensions, and based on these results we may partition $ {\bd g}$ as follows\footnote{General third order tensors include additional terms
${\bd g}^{(0)}$ and ${\bd g}^{(2)}$ which vanish for a symmetric tensor
\citep{Jerphagnon70}.}
\beq{401}
 {\bd g} =  {\bd g}^{(1)} + {\bd g}^{(3)},
 \eeq
 where
  \beq{402}
 g^{(1)}_{\alpha \beta \gamma}
 =  \frac14 \big( t_\alpha \delta_{ \beta \gamma} +t_\beta\delta_{  \gamma\alpha} +t_\gamma \delta_{\alpha \beta }\big) ,
 \qquad
 t_\alpha = g_{\alpha \beta \beta}.
 \eeq
 ${\bd g}^{(1)}$ is pseudovector \citep{Jerphagnon70} with two independent components,
 $(g_{122}+g_{111})$ and  $(g_{112}+g_{222})$.
  ${\bd g}^{(3)}$, which also has two independent elements,
   $(3g_{122}-g_{111})$ and  $(3g_{112}-g_{222})$,
   may be called a dimer.  More importantly, it is harmonic, i.e.
 $g^{(3)}_{\alpha \beta\beta } =0$.  Hence, the  quadratic invariant
 $g_{\alpha \beta \gamma} g_{\alpha \beta \gamma}$ can be expressed
 \beq{044}
  g_{\alpha \beta \gamma} g_{\alpha \beta \gamma}
 =   \gamma_1 + \gamma_3 ,
 \eeq
where $\gamma_1$ and $\gamma_2$ are the quadratic invariants of the constituent tensors:
 \begin{align}\label{045}
 \gamma_1
 &=  %\frac34 |{\bd t}|^2 =
  g_{\alpha \beta \gamma}^{(1)} g_{\alpha \beta \gamma}^{(1)} =
 \frac34 \big(g_{111}+g_{122}\big)^2
 + \frac34\big(g_{112}+g_{222}\big)^2,
 \\
 \gamma_3
 &=  g_{\alpha \beta \gamma}^{(3)} g_{\alpha \beta \gamma}^{(3)}=
 \frac14  \big(3g_{122}- g_{111}\big)^2
 + \frac14 \big(3g_{112}-g_{222}\big)^2.
 \end{align}
 Comparing these with eq. \rf{27} indicates that
 $\mu = \frac12 \gamma_3 - \frac16 \gamma_1$, and hence the no coupling condition can be expressed in terms of  invariants as { $\gamma_1  = 3\gamma_3$.
 Alternatively, noting that $\gamma_1  = \frac34 g_{\alpha \beta\beta}g_{\alpha \gamma\gamma}$, we deduce
 $\mu = \frac12 g_{\alpha \beta\gamma}g_{\alpha \beta\gamma}-
 \frac12 g_{\alpha \beta\beta}g_{\alpha \gamma\gamma}$, and hence
 \beq{271}
\mbox{no coupling}\quad \Leftrightarrow \quad
g_{\alpha \beta\gamma}g_{\alpha \beta\gamma}=
g_{\alpha \beta\beta}g_{\alpha \gamma\gamma}.
\eeq}

{
\begin{examp}
We check the condition eq. \rf{-6} in two cases of acoustic axes:
$[1 \, 0 \, 0]$ and $[1 \, 1 \, 1]$ for the cubic crystal considered
earlier. It is easy to see that the
condition \rf{-6} is satisfied and the evolution equations for shear
waves are decoupled for the $[1 \, 0 \, 0]$ acoustic axis. However, for the $[1 \, 1 \, 1]$ axis, the left hand
side of eq. \rf{-6} is equal to $-2 \Gamma_{s + 2} \neq 0$, so the
condition \rf{-6} is not satisfied. Therefore there is quadratically nonlinear  coupling in the evolution equations for shear waves in this case (see eqs.
\rf{83} and \rf{63}).
\end{examp}}

\section{Concluding remarks}

Starting from a formulation  of the governing equations as a first
order system of quasilinear equations, we have derived the general
form of the amplitude evolution equations for weakly nonlinear plane
wave propagation.   The major new results concern the
form of the  evolution equations for the degenerate conditions
associated with propagation along acoustic axes, summarized in
Lemma 1 and eqs. \rf{-2}.  The quasi-transverse wave amplitudes are
coupled at the quadratically nonlinear level, with at most four
interaction coefficients.  The number of coefficients reduces in the
presence of symmetry, with the precise number determined by Lemma 2.
For instance, the coupling in the presence of three-fold symmetry
about the propagation direction depends on  a single interaction
coefficient, with the canonical form of the coupled equations given
by eqs. \rf{83}.  The nonlinear coupling disappears if the acoustic
axis has four-fold or higher symmetry. The isotropic solid is of
course the most obvious example, but the results presented here show
that similar decoupling can be  expected in the presence of
anisotropy. We have also shown that it is possible for the coupling
to vanish even when the acoustic axis is
not a symmetry axis.   The condition, defined by Lemma 3, requires
that the interaction coefficients satisfy a unique relation.  Taken
together, the variety of results presented here shed light on the
nature of the equations governing nonlinear wave propagation in
elastic crystals.

\section*{Acknowledgment}
W.D. acknowledges partial support from the Polish State Committee
grant No. 0 T00A 014 29.

%%%%%%%%%%%%%%%%%%%%%%%%%%%%%%%%%%%%%%%%%%
\appendix

\section{Appendix: Cubic crystals}\label{sec7}

In a cubic crystal with cube axes ${\bd e}_j$, $j=1,2,3$, we have (see
\cite{Domanski08b})
\begin{align*}
c_{abcd}E_{ab}E_{cd} &= c_{11}(I_1^2 - 2I_2 ) +2c_{12} I_2 + 4c_{44}I_3,
\\
c_{abcdef}E_{ab}E_{cd}E_{ef}&=
c_{111}(I_1^3 - 3 I_1I_2+3I_4)
+3 c_{112}(  I_1I_2-3I_4)
\\ & \quad
+12 c_{144}(  I_1I_3- I_6)
+6c_{123}I_4
+48c_{456}I_5
+12c_{166}I_6,
\end{align*}
where ${\bd E}^t = {\bd E}$ and
\begin{align*}
I_1 &= E_{11}+E_{22}+E_{33},
\\
I_2 &= E_{11}E_{22}+E_{22}E_{33}+E_{33}E_{11},
\\
I_3 &= E_{12}^2+E_{23}^2+E_{31}^2,
\\
I_4 &= E_{11} E_{22} E_{33},
\\
I_5 &= E_{12}E_{23}E_{31},
\\
I_6 &= (E_{11}+E_{22})E_{12}^2+(E_{22}+E_{33})E_{23}^2+(E_{33}+E_{11})E_{31}^2.
\end{align*}

The four cube diagonals are axes of trigonal or 3-fold symmetry, and are acoustic axes.
Consider  propagation along the cube diagonal acoustic axis $\bd {n} = [1,1,1]/\sqrt{3}$, and assume $s=1$.  The parameters $\Gamma_s$  and  $\Gamma_{s+2}$  follow by taking
${\bd k}_1$ orthogonal to one of the 3 symmetry planes, e.g.
${\bd k}_1 = [1,-1,0]/\sqrt{2}$ and
${\bd k}_2 = [-1,-1,2]/\sqrt{6}$.  The coefficient  $\Gamma_1$ is then obtained by using
\begin{align*}
{\bd E} = \frac12({\bd n}\otimes {\bd k}_1 + {\bd k}_1 \otimes {\bd n})
= \frac1{2\sqrt{6}}
\begin{bmatrix}
2 & 0 & 1 \\
0 & -2 & -1 \\
1& -1 & 0
\end{bmatrix}.
\end{align*}
Therefore,
$I_1=I_4=I_5=I_6=0$ implying   $c_{abcdef}E_{ab}E_{cd}E_{ef} =0$ and hence
$\Gamma_s=0$,  as expected for a two-fold symmetry axis.    The coefficient $\Gamma_3
=\Gamma_{s+2}$ follows from eq. \rf{51}
with
\begin{align*}
{\bd E} = \frac12({\bd n}\otimes {\bd k}_2 + {\bd k}_2 \otimes {\bd n})
= \frac1{6\sqrt{2}}
\begin{bmatrix}
-2 & -2 & 1 \\
-2 & -2 & 1 \\
1& 1 & 4
\end{bmatrix},
\end{align*}
leading to
\begin{align*}
c_{abcdef}E_{ab}E_{cd}E_{ef}
= \frac{1}{9\sqrt{2}  }
 \big[
 c_{111} +2c_{123}-2 c_{456}  -3(c_{112}-c_{144}+c_{166 })
 \big].
\end{align*}

%%%%%%%%%%%%%%%%%%%%%%%%%%%%%%%%%%%%%%%%%%%
%\bibliography{Domanski}
%\bibliographystyle{plainnat}%unsrtnat}%abbrvnat}%plain}%doipubmed}%harvard}%unsrt}%natbib}

\end{document}